\documentstyle[12pt]{article}
\newcommand{\bce}{\begin{center}}
\newcommand{\ece}{\end{center}}
\newcommand{\beq}{\begin{equation}}
\newcommand{\eeq}{\end{equation}}
\newcommand{\bea}{\vspace{0.25cm}\begin{eqnarray}}
\newcommand{\eea}{\end{eqnarray}}

\newcommand{\ba}{\begin{array}}
\newcommand{\ea}{\end{array}}

\newcommand{\doublespace}{
    \renewcommand{\baselinestretch}{1.6}\large\normalsize}

\def\lsim{\mathrel{\rlap{\lower4pt\hbox{\hskip1pt$\sim$}}
    \raise1pt\hbox{$<$}}}     
\def\gsim{\mathrel{\rlap{\lower4pt\hbox{\hskip1pt$\sim$}}
    \raise1pt\hbox{$>$}}}     

\def\lsim{\mathrel{\rlap{\lower4pt\hbox{\hskip1pt$\sim$}}
    \raise1pt\hbox{$<$}}}         
\def\gsim{\mathrel{\rlap{\lower4pt\hbox{\hskip1pt$\sim$}}
    \raise1pt\hbox{$>$}}}         

\def\beq{\begin{equation}}
\def\endeq{\end{equation}}
\def\arr{\begin{eqnarray}}
\def\endarr{\end{eqnarray}}

\makeindex
\begin{document}
\phantom{.}\hspace{8.5cm}{preprint KFA-IKP(TH)-1994-29} \\
\phantom{.}\hspace{8.0cm}{8.August 1994}
\vspace{2cm}
\begin{center}
{\bf \huge Simple classification of final state interaction
effects in $^{4}He(e,e'p)$ scattering \\}
\vspace{1cm}
{\bf A.Bianconi$^{1)}$, S.Jeschonnek$^{2)}$,
N.N.Nikolaev$^{2,3)}$, B.G.Zakharov$^{3)}$ } \medskip\\
{\small \sl
$^{1)}$Dipartimento di Fisica Nucleare e Teorica, Universit\`a
di Pavia, and\\
Istituto Nazionale di Fisica Nucleare,
Sezione di Pavia, Pavia, Italy \\
$^{2)}$IKP(Theorie), Forschungszentrum  J\"ulich GmbH.,\\
D-52425 J\"ulich, Germany \\
$^{3)}$L.D.Landau Institute for Theoretical Physics, \\
GSP-1, 117940, ul.Kosygina 2, V-334 Moscow, Russia
\vspace{1cm}\\}
{\bf \LARGE A b s t r a c t \bigskip\\}
\end{center}
The radius of interaction
between the struck proton and spectator nucleons is close to the
radius of short-distance two-nucleon correlations in nuclear
matter, which makes final state interaction (FSI) an
important background to production of protons with large missing
momentum. We present a simple classification of the dominant FSI
effects in $^{4}He(e,e'p)$ scattering and identify parts of the
phase space dominated by FSI. At large missing momentum, final
state interaction leads to a striking angular anisotropy of
the missing momentum distribution, which has a prominent peak
in transverse kinematics and smaller, forward-backward
asymmetric, peaks in parallel kinematics.
\medskip\\
{\bf PACS: 25.30Fj,~24.10Eq}

\newpage
\doublespace

Quasielastic $(e,e'p)$ scattering on nuclei at large missing
momentum $p_{m}$ is a relevant tool for investigation of
short-distance nucleon-nucleon interaction (initial state
two-nucleon correlations (ISC)) in the nuclear medium [1] and
constitutes an important part of the CEBAF experimental
program [2]. The
measured $p_{m}$ distribution is distorted by
final state interaction (FSI) of
the struck proton in the target nucleus debris. The
point we wish to make in this paper is that, in the CEBAF
range of energies,
the radius $b_{o}$ of the FSI is very close to the
radius $r_{c}$ of ISC (see below).
This makes FSI an important background
in the production of protons with large $p_{m}$, which may obscure the
relationship between the observed $p_{m}$
distribution and the ISC. Indeed,
a strong effect of FSI in $(e,e'p)$ scattering on  $C$ and $Pb$
nuclei was found in [3].

In this note we present simple estimates of the effect
of FSI in $^{4}He(e,e'p)$ scattering. We show how FSI
leads to an anisotropic $p_{m}$
distribution, with the dominance of FSI effects in transverse
kinematics and significant FSI corrections to
ISC effects in parallel kinematics. This last point can
be important for the $y$-scaling analysis.
We find a novel effect of
quantum-mechanical interference of ISC and FSI
which gives a large correction to the
elastic rescattering of the struck proton
and which must be included when discussing
more sophisticated features of FSI such as color
transparency effects [4].
The emphasis of this paper is on the classification of FSI effects
and on semi-analytic estimates of
the relative magnitude of FSI and ISC effects. We
concentrate on the region of 4-momentum transfer squared $Q^{2}
\gsim $(1-2)GeV$^{2}$, which is relevant to the planned CEBAF
experiments [2]. Furthermore, in this range of $Q^{2}$, FSI can
be described by Glauber's multiple scattering theory [5], which
simplifies the evaluation of FSI effects.

The reduced nuclear amplitude for the exclusive process
$^{4}He(e,e'p)A_{f}$ is given by
\arr
{\cal M}_{f}=
\int d\vec{R}_{1}\,d\vec{R}_{2}\,d\vec{R}_{3}
\Psi_{f}^{*}(\vec{R}_{1},\vec{R}_{2})
S(\vec{r}_{1},...,\vec{r}_{4})
\Psi(\vec{R}_{1},\vec{R}_{2},\vec{R}_{3})
\exp(i\vec{p}_{m}\vec{R}_{3})
\label{eq:1}
\endarr
where $\Psi(\vec{R}_{1},\vec{R}_{2},\vec{R}_{3})$ and $
\Psi_{f}(\vec{R}_{1},\vec{R}_{2})$ are wave functions of
the target $^{4}He$ nucleus and of the specific 3-body final state
$A_{f}$, which are conveniently  described in terms of the
Jacobi coordinates $\vec{R}_{1}=\vec{r}_{2}-\vec{r}_{1}$,
{}~$\vec{R}_{2}={2\over 3}\vec{r}_{3}-{1\over 3}(\vec{r}_{1}+
\vec{r}_{2})$,~$\vec{R}_{3}=\vec{r}_{4}-{1\over 3}(\vec{r}_{1}+
\vec{r}_{2}+\vec{r}_{3})$ (plus $\vec R_{cm}  = {1\over 4}
\sum_i \vec r_i
 \equiv  0$). Lab coordinates
$\{\vec r_i(\vec R_j,\vec R_{cm})\}$
are also used in the text for sake of intuition.
The nucleon ``4'' of $^{4}He$ is
chosen for the detected struck proton with momentum $\vec{p}$,~
$\vec{p}_m$ $\equiv$ $\vec q - \vec p$ is the missing momentum,
$\vec{q}$ is the $(e,e')$ momentum transfer
and
$S(\vec{r}_{1},...,\vec{r}_{4})$ is the $S$-matrix of the
FSI of the struck proton with three spectator nucleons.
In this paper we discuss the quantity $\sum_{f}|M_{f}|^{2}$, which
gives the inclusive spectrum of protons in $^{4}He(e',e)$ scattering
in quasielastic kinematics. Summing over all the allowed final
states $A_{f}$
for the three undetected nucleons, the closure relation
\beq
\sum_{f}
\Psi_{f}(\vec{R}_{1}\,',\vec{R}_{2}\,')
\Psi_{f}^{*}(\vec{R}_{1},\vec{R}_{2}) =
\delta(\vec{R}_{1}-\vec{R}_{1}\,')
\delta(\vec{R}_{2}-\vec{R}_{2}\,')\,
\label{eq:2}
\endeq
leads to \footnote{The related formula of Ref.[6] introduces
an extraneous factor ${3\over 4}$ in the exponent of the Fourier
transform. A
consistent treatment of the center of mass motion of the 3-body
final state leads to our Eq.~(3).}
\arr
\label{eq:3}
w(\vec{p}_{m}) =
\sum_{f} |{\cal M}_{f}|^{2} =
\int d\vec{R}_{1}\,d\vec{R}_{2}
 d\vec{R}_{3}\,'d\vec{R}_{3}
\exp\left[i\vec{p}_{m}(\vec{R}_{3}-\vec{R}_{3}\,')\right]~~~~~~~~~~
\nonumber\\
\times \Psi^{*}(\vec{R}_{1},\vec{R}_{2},\vec{R}_{3}\,')
S^{\dagger}(\vec{r}_{1}\,'..,\vec{r}_{4}\,')
S(\vec{r}_{1},...,\vec{r}_{4})
\Psi(\vec{R}_{1},\vec{R}_{2},\vec{R}_{3})\, .
\endarr
In the absence of FSI, $w(\vec{p}_{m})$ coincides with the familiar
single-particle momentum distribution $n_{F}(p_{m})$, extensive
studies of which are available in the literature [7].
High-$\vec{p}_{m}$ Fourier
components in Eq.~(3) come from the rapid variation of
the integrand with $|\vec{R}_{3}-\vec{R}_{3}'|$, which originates
from either ISC or FSI.
In the following, ``long ranged'' will indicate all the functions
which change on the scale of the
$^{4}He$ radius $R_{o}\approx
1.4 fm$ ([8,9], for more accurate definition see below),
while ``short
ranged'' refers to changes on the scale of the correlation
radius $r_{c}
\sim 0.5 fm$ [10] and/or the FSI radius $b_{0}$.
A simple way of modeling the ISC
effect is
\beq
\Psi(\vec{R}_{1},\vec{R}_{2},\vec{R}_{3})\equiv\
\Psi_{o}(\vec{R}_{1},\vec{R}_{2},\vec{R}_{3}) F, ~~{\rm where}~~
F \equiv\
\prod_{i<j}^{4}\Big[1-C(\vec{r}_{i}-\vec{r}_{j})\Big]\, .
\label{eq:4}
\endeq

Here $\Psi_{o}$ is a (long range) mean field wavefunction,
and
$C(\vec{r})$ is a short range correlation function.
For a hard core repulsion
$C_{o}=C(0)=1$, for a soft core $C_{o} < 1$.
At the large $Q^{2}$ of interest in the CEBAF experiments, FSI
can be described by Glauber theory.
Defining transverse and longitudinal components
$\vec{r}_{i}$ $\equiv$
$(\vec{b}_{i}+z_{i}\hat q)$ and $\vec{R}_{i}$ $\equiv$
$(\vec{B}_{i}+Z_{i}\hat q)$ we can write
\beq
S(\vec{r}_{1},...,\vec{r}_{4}) = \prod_{i=1}^{3}
\Big[ 1-\theta(z_{i}-z_{4})\Gamma(\vec{b}_{4}-\vec{b}_{i})\Big],
\label{eq:5}
\endeq
where $\Gamma(\vec{b})$ is the profile function of the
nucleon-nucleon scattering
\beq
\Gamma(\vec{b})
\ \equiv\
{ \sigma_{tot} (1 - i \rho) \over 4 \pi b_{o}^2  }
\exp \Big[-{\vec{b}^2 \over 2 b_{o}^2} \Big]
\label{eq:6}
\endeq
($\rho$ is the ratio of the real to imaginary part of the
forward elastic scattering amplitude). The
Glauber formalism describes quite well nucleon-nucleus
scattering at energies from 500 MeV to many GeV, even at angles
as large as 30$^o$ at 500 MeV (for a review
see [11]). At $T_{kin} \sim 1$GeV
$b_{o}\approx 0.5 fm$ and $\sigma_{tot} \approx
40$mb [11,12,13].

Because of $r_{c}^{2},b_{o}^{2} \ll R_{o}^{2}$, the rapid variation
of the integrand of $w(\vec{p}_{m})$ comes from
\arr
FF^+SS^+\ =\
\prod_{i<j}^{4}\Big[1-C(\vec{r}_{i}\,'-\vec{r}_{j}\,')\Big]
\Big[1-C(\vec{r}_{i}-\vec{r}_{j})\Big] \nonumber \\
\times
\prod_{i\neq 4}
\Big[ 1-\theta(z_{i}'-z_{4}')\Gamma^{*}
(\vec{b}_{4}\,'-\vec{b}_{i}\,')\Big]
\Big[ 1-\theta(z_{i}-z_{4})\Gamma(\vec{b}_{4}-\vec{b}_{i})\Big]=
\nonumber\\
1-\sum_{i<j}\left[C'+C\right]
-\sum_{i\neq 4}
\left[\Gamma' +
\Gamma \right]
+ \sum [C'\Gamma +C\Gamma'] + \sum C'C + \sum \Gamma' \Gamma +....
\label{eq:7}
\endarr
The inequalities $r_{c}^{2},b_{o}^{2} \ll R_{o}^{2}$ allow
to develop a simple
classification of ISC and FSI effects, which have different
$\vec{p}_{m}$ dependence and absolute normalization.
It is convenient to demonstrate these properties upon the
"exactly soluble" model with the harmonic oscillator (HO)
mean field wave function
\beq
\Psi_{o} \propto
\exp \Big[-{1\over 2R_{o}^{2}}\sum_{i}^{4} \vec{r}_{i}\,^{2} \Big]
=
\exp \Big[-{1\over 4R_{o}^{2}}\left(\vec{R}_{1}\,^{2}+
3\vec{R}_{2}\,^{2}+{3\over 2}\vec{R}_{3}\,^{2}\right)
\Big]
\label{eq:8}
\endeq
and the simple parameterization for the correlation function
$
C(r)=C_{o}\exp\left(-{r^{2}/2r_{c}^{2}}\right).
$

First, let us comment briefly on ISC effects neglecting FSI.
To the zeroth order in $C(\vec{r}_{i}-\vec{r}_{j})$, the correlations
are not present at all.
The resulting $p_{m}$
distribution vanishes rapidly at $p_{m} \gsim k_{F} \sim 1/R_{o}$.
In the HO model (8) one finds
\beq
w(1;\vec{p}_{m})= w_{1}\exp\Big(-{4\over 3}R_{o}^{2}p_m^2\Big).
\label{eq:9}
\endeq
($w(1;\vec{p}_{m})$, $w(C;\vec{p}_{m})$,... indicate
the contributions to
$w(\vec{p}_{m})$ coming from the ``1'',``C'',... terms
in eq.(7)).

Analogous contributions (but smaller in magnitude by the factor
$\propto (r_{c}/R_{0})^{3}$)
come from the terms which are
linear in $C(\vec{r}_{i}-\vec{r}_{j})$:
their short range
behaviour is averaged away by the integrations in $dR_1 dR_2$.
With the choice (8) one  finds
\beq
w(C;\vec{p}_{m})\approx -w_{1} C_o 4 \sqrt{\frac {243} {125}}
\Big({r_c \over R_o}\Big)^3
exp\Big[-{4 \over 5}R_{o}^{2}p_{m}^{2}\Big].
\label{eq:10}
\endeq
For the sake of brevity, we don't show here correction
factors $[1+{\cal O}(r_{c}^{2}/R_{o}^{2}) ]$ to slopes and
normalization factors in eq. (10),(11),(13)-(20).
They will be presented elsewhere [14]. They are anyway
included in exact form in all the numerical results to be
presented below.

The large-$p_{m}$ component of the momentum distribution
comes from the three identical terms of the form $C(\vec{r}_{4}-
\vec{r}_{i})C(\vec{r}_{4}\,'-\vec{r}_{i}\,')$.
The corresponding contribution $w(C'C;\vec{p}_{m})$
to $w(\vec{p}_{m})$ directly probes $C(\vec{r})$:
\arr
w(C'C;\vec{p}_{m}) \approx w_{1}
\sqrt {\frac{243}{512}} {1\over R_{0}^{6}}
{ 1 \over (2 \pi)^3}
\left|\int d^{3}\vec{R}_{3}
C(\vec{R}_{3}) \exp(i\vec{p}_{m}\vec{R}_{3})\right|^{2} \nonumber\\
\approx
w_{1} C_{o}^{2} \sqrt {\frac{243}{512}}
\left ({r_c \over R_o} \right)^{6}
\exp\left(-r_{c}^{2}
p_{m}^{2}\right)
\label{eq:11}
\endarr
Notice the small normalization factor $\propto
(r_{c}/R_{o})^{6}$.
With $r_{c} = 0.5$fm, $C_{o} = 1$
 and $R_{o}=1.4fm$, the above estimated $w(C'C;\vec{p}_{m})$
is in good agreement with the tail of $n_{F}(p_{m})$ as given by
Ciofi degli Atti $et$ $al$ [7].
All the above discussed
terms give an isotropic $\vec{p}_{m}$ distribution.

The classification of FSI effects is very similar to the above
with one important difference: $C(\vec{r}_{4}-
\vec{r}_{i})$ is a short-ranged isotropic function of
$\vert \vec r_4-\vec r_i\vert$, whereas
$\Gamma(\vec{b}_{4}-\vec{b}_{i})\theta
(z_{i}-z_{4})$ is a short-ranged function of the transverse
separation $|\vec{b}_{4}-\vec{b}_{i}|$ only. In the longitudinal
direction the FSI operator behaves as
a long-ranged one, which
leads to the angular anisotropy of FSI
effects. We decompose $\vec{p}_{m}\ \equiv\ (\vec{p}_{\perp}+
p_{m,z}\hat{q})$. Evidently, the $p_{\perp}$ dependence of terms
linear in $\Gamma(\vec{b}_{4}-\vec{b}_{i})$ is the same as in
Eq.(10) with $r_{c}$ substituted for $b_{0}$. The
dependence on $p_{m,z}$ will be more similar to that in Eq.(9),
apart from distortions at large $p_{m,z}$ coming
from the high frequency Fourier components of
$\theta(z_{i}-z_{4})$.
Furthermore, the finite real part of the forward $pN$ scattering
amplitude leads to a forward-backward asymmetry of the $p_{m,z}$
distribution [15]. The distortion and asymmetry effects
(which can be clearly seen in the figures 1 and 2) will be
discussed in more detail elsewhere [14]. For each $\Gamma$
we have the normalization
\beq
Y={b_{o}^{2} \over R_{o}^{2}}
\cdot{\sigma_{tot} \over 4\pi b_{o}^{2}}
={\sigma_{tot}\over 4\pi R_{o}^{2}} \approx 0.17
\label{eq:12}
\endeq
(see eq.(6); $b_o^2/R_o^2$ comes from the integration)
which is larger than the normalization in (10) by a factor
$\sim (R_o/r_{c})$. The linear FSI terms give negative-valued
contributions
to $w(\vec p_m)$, expressing the direct reduction of the
proton flux by FSI.
At large $p_{\perp}$, the leading FSI
component of $w(\vec{p}_{m})$ comes from the terms $\propto \Gamma
(\vec{b}_{4}-\vec{b}_{i})\Gamma^{*}(\vec{b}_{4}\,'-\vec{b}_{i}\,')$,
which describe the elastic rescattering of the struck proton
on the spectator nucleon "i":
\arr
w(\Gamma'\Gamma;\vec{p}_{m}) \propto
\left|\int d^{2}\vec{B}_{3}
\Gamma(\vec{B}_{3})
\exp(i\vec{p}_{\perp}\vec{B}_{3})\right|^{2}
=
4\pi {d\sigma_{el} \over dp_{\perp}^{2}} =
{1 \over 4}\sigma_{tot}^{2}(1+\rho^{2})\exp(-b_{o}^{2}p_{\perp}^{2})
\label{eq:13}
\endarr
 Because of $b_{o}\approx r_{c}$ in
the CEBAF range of $Q^{2}$, the $w(C'C;\vec{p}_{m})$ and
$w(\Gamma'\Gamma;\vec{p}_{m})$ components have similar $p_{\perp}$
dependence. However (compare eqs. (10) and (12)) the overall
normalization is larger for the FSI term. So
at $p_{m,z}=0$ we find a strong dominance of the FSI
rescattering effect over the ISC effect:
\beq
{\Gamma'\Gamma \over
C'C} \approx
{1 \over C_{o}^{2}\sqrt{6}}\cdot
\left[{\sigma_{tot} \over 4\pi r_{c}^{2}}\right]\cdot
{\left(R_{o} \over r_{c}\right)^{2}} \sim 7.
\label{eq:14}
\endeq
Because the FSI are long-ranged in $z_4-z_i$,
$w(\Gamma'\Gamma;\vec{p}_{m})$ decreases
steeply with $p_{m,z}$ on the scale $\sim k_{F}$, which
leads to
an angular anisotropy of the elastic rescattering effect.
The effect of quantum-mechanical interference of
ISC and FSI  comes from the terms $\propto C(\vec{r}_{4}
    \,'-
\vec{r}_{i}\,')\Gamma(\vec{b}_{4}-\vec{b}_{i})$ and $\propto
C(\vec{r}_{4}-\vec{r}_{i})\Gamma^{*}(\vec{b}_{4}\,'-\vec{b}_{i}\,')$.
The corresponding contribution $w(C\Gamma'+C'\Gamma;\vec{p}_{m})$
to $w(\vec{p}_{m})$ has the $p_{\perp}$ dependence
\beq
w(C\Gamma'+C'\Gamma;\vec{p}_{m})\propto
\exp\left[-{1\over 2}(r_{c}^{2}+b_{o}^{2})p_{\perp}^{2}\right]\, ,
\label{eq:17}
\endeq
steep $p_{m,z}$ dependence similar to that in Eq.(10), and the
large normalization
\beq
{C\Gamma'+C'\Gamma \over \Gamma'\Gamma} = 4 \sqrt{{3\over 5}}
C_{o}\left(4\pi r_{c}^{2} \over \sigma_{tot}\right)
\cdot {r_{c}\over R_{o}} \sim 1\,.
\label{eq:18}
\endeq
Notice that any quasiclassical
consideration would completely miss this large ISC-FSI
interference effect.

The leading ISC correction to the elastic rescattering effect
comes from the terms $\propto C(\vec{r}_{4}-\vec{r}_{i})\Gamma
(\vec{b}_{4}-\vec{b}_{i})\Gamma^{*}(\vec{b}_{4}\,'-\vec{b}_{i}\,')$,
which have the $p_{\perp}$ dependence with the slope
\beq
{1\over 2}
\left(b_{o}^{2}+ {b_{o}^{2}r_{c}^{2} \over b_{o}^{2}+r_{c}^{2}}
\right) < b_{o}^{2} \, ,
\label{eq:15}
\endeq
a steep $p_{m,z}$-dependence (as in Eq.(11)) and the
normalization
\beq
{C\Gamma'\Gamma + C'\Gamma'\Gamma \over \Gamma'\Gamma} =
-2C_{o}\sqrt{{3\over 5}}\cdot{r_{c}^{2} \over b_{o}^{2} + r_{c}^{2}}
\cdot {r_{c} \over  R_{o}} \sim -0.3.
\label{eq:16}
\endeq
These corrections are not small.
In a different form, a similar result is contained
in Eq.~(7) of Ref. [3].
The FSI correction to the ISC effect comes from terms
$\propto C(\vec{r}_{4}\,'-\vec{r}_{i}\,')C(\vec{r}_{4}-
\vec{r}_{i}) \Gamma(\vec{b}_{4}-\vec{b}_{i})$, which give
\beq
w(C'C\Gamma+C'C\Gamma';\vec{p}_{m}) \propto
\exp(-r_{c}^{2}p_{m,z}^{2})
\exp\left[-
{1\over 2}\left({b_{o}^{2}r_{c}^{2} \over b_{o}^{2}+r_{c}^{2}}
+r_{c}^{2}\right)p_{\perp}^{2}\right],
\label{eq:19}
\endeq
slow decrease with $p_{z}$ and the relative normalization
\beq
{C'C\Gamma+C'C\Gamma'\over  C'C} \approx -
{r_{c}^{2} \over r_{c}^{2}+b_{o}^{2}}\cdot
{\sigma_{tot} \over 4\pi r_{c}^{2}} \sim -{2\over 3} \, .
\label{eq:20}
\endeq
This is reminiscent of the large probability of double-scattering
in a small deuteron-like 2-particle cluster
of size $\sim r_{c}$ [5].
One can
evaluate the still higher-order terms
$\propto C(\vec{r}_{4}\,'-\vec{r}_{i}\,')
C(\vec{r}_{4}-\vec{r}_{i}) \Gamma(\vec{b}_{4}-\vec{b}_{i})
\Gamma^{*}(\vec{b}_{4}\,'-\vec{b}_{i}\,')$, which have a broad
$(p_{\perp},p_{m,z})$-distribution, but small absolute normalization
[14].

In Fig.~1 we show the angular dependence of $w(\vec{p}_{m})$
and its most important components  at large $p_{m}$.
The large-$p_{m}$ tail of the undistorted distribution
$n_{F}(p_{m})$ is borrowed from ref. [7]. For the
elastic rescattering
$w(\Gamma'\Gamma;\vec{p}_{m})$ and the ISC-FSI
interference $w(C\Gamma'+C'\Gamma;\vec{p}_{m})$
we take our results (shown in approximated form in eqs.
(13),(14) and (17),(18), respectively). We find a prominent
signal of the elastic rescattering and ISC-FSI interference
in parallel kinematics.
Notice a significant
forward-backward asymmetry of the ISC-FSI
interference component $w(C\Gamma'+C'\Gamma;\vec{p}_{m})$, which
is generated by the real part of the $pN$ scattering amplitude,
and also the unexpectedly large contribution from the elastic
rescattering term at $p_{\perp}=0$,
which comes from the
$\theta$-function generated tail of the $p_{z}$ distributions.
To illustrate this general FSI effect,
in Fig.~2 we show what happens if
one substitutes ${1\over 4}$ for the product of
$\theta(z_{i}-z_{4})$ and $\theta(z_{i}-z_{4}')$. The distortion
effect becomes stronger at larger $p_{m}$, leading to the forward
and backward peaks alongside the FSI peak in the transverse
kinematics. This could influence the $y$-scaling analysis (see
also [19]).

Summarizing the main results, we have presented a simple
classification of FSI effects compared to the ISC effects.
The anisotropic angular dependence of
the elastic rescattering and ISC-FSI interference effects
compared to the isotropic momentum distribution in the PWIA is
a striking signature of FSI effects.
An important finding
is that apart from the peak at $90^{o}$,
distortions by FSI lead also to forward and
backward peaks, which slowly
build up with increasing missing momentum $p_{m}$.
We found a large contribution to
the $p_{m}$ distribution in transverse kinematics from
the novel effect of the quantum mechanical interference of
FSI and ISC effects.
The considered aspects of ISC and FSI were not discussed in
previous work on FSI and ISC effects in $A(e,e'p)$ scattering
[16,17,18].
Our results suggest that the experimental separation of the
FSI and ISC effects is more difficult than thought before.
The expansion parameter of the problem
$\sim r_{c}/R_{o}$ is not very small, and more detailed numerical
analysis with allowance for the tensor correlation function
and the D-wave in $^{4}He$ are called upon to establish the
sensitivity to models for the correlation function.
The implications of the attenuation, distortion and
forward-backward asymmetry effects for the $y$-scaling analysis
in terms of the undistorted ISC effect will be presented
elsewhere.

{\bf Acknowledgements:} We acknowledge discussions with
S.Boffi and J. Speth. A. Bianconi thanks
C.Ciofi
degli Atti and S.Simula for discussions
on the D-wave and tensor correlation
effects.
This work was supported in part by the
INTAS Grant No. 93-239 and by the Vigoni Program of DAAD (Germany)
and the Conferenza Permanente dei Rettori (Italy).
A.Bianconi thanks J.Speth for the hospitality at IKP, KFA J\"ulich,
and S.Jeschonnek thanks S.Boffi for the hospitality at the
University of Pavia.

\pagebreak

{\bf \Large Figure captions:}

The figures can be obtained from the authors via e-mail:

KPH127@DJUKFA11\\

or \\

s.jeschonnek@kfa-juelich.de \\

\begin{itemize}

\item[{\bf Fig.~1}]
   The angular dependence of the three most important components
   of $w(\vec{p}_{m})$ at $p_{m}= 2 fm^{-1}$ (lower panel)
   and at $p_{m}= 3 fm^{-1}$ (upper panel).
   The dotted line is
   the undistorted distribution $n_{F}(p_{m})$ taken from
   \cite{7}, the dash-dotted line represents the ISC-FSI
   interference component $w(C\Gamma'+C'\Gamma;\vec{p}_{m})$,
   the dashed line shows the elastic rescattering component
   $w(\Gamma'\Gamma;\vec{p}_{m})$, and the solid line is the sum
   of the three contributions.

\item[{\bf Fig.~2}]
   The angular dependence of the elastic rescattering term
   $w(\Gamma'\Gamma;\vec{p}_{m})$
   including the product $\theta(z_{i}-z_{4})$ and
   $\theta(z_{i}'-z_{4})$ (solid line) and replacing the product of the
   $\theta$ -functions by $\frac {1}{4}$ (dotted line) for
   $p_m = 2 fm ^{-1}$ (lower panel) and $p_m = 3 fm ^{-1}$
   (upper panel).

\end{itemize}

\begin{thebibliography}{299}

\bibitem{1} 
A.E.I.Dieperink and P.K.A. de Witt Huberts,
{\sl Annu. Rev. Nucl. Part. Sci.} {\bf 40} (1990) 239;
S.Boffi, C.Giusti and F.D.Pacati,
{\sl Phys.Rep.} {\bf 226} (1993) 1. These reviews contain
extensive list of references to early work on DWIA description
of $(e,e'p)$ scattering.

\bibitem{2} 
J.Mougey (spokesperson), CEBAF Proposal No. E-89-044;
R.G.Milner (spokesperson), CEBAF Proposal No. E-91-007;
D.F.Geesaman (spokesperson), CEBAF Proposal No. E-91-011.

\bibitem{3} 
N.N.Nikolaev, A.Szczurek, J.Speth, J.Wambach, B.G.Zakharov
and V.R.Zoller,
J\"ulich preprint {\bf KFA-IKP(Th)-1993-31} (1993), submitted
to {\sl Nucl. Phys.} {\bf A}.

\bibitem{4} 
N.N.Nikolaev,
{\sl Surveys in High Energy Physics} {\bf v.7, No.1-4} (1994) pp.1-92,
Harwood Academic Publishers, Basel;
N.N.Nikolaev and B.G.Zakharov, Color transparency after the NE18 and
E665 experiments: Outlook and perspectives at CEBAF, {\sl J\"ulich
preprint} {\bf KFA-IKP(Th)-1994-20}, to be published in Proceedings
of the Workshop on CEBAF at Higher Energies, CEBAF, 14-16 April 1994.

\bibitem{5} 
R.J.Glauber, in: {\sl Lectures in Theoretical Physics}, v.1,
ed. W.Brittain and L.G.Dunham. Interscience Publ., N.Y., 1959;
R.J.Glauber and G.Matthiae, {\sl Nucl. Phys.} {\bf B21} (1970) 135.

\bibitem{6} 
H.Morita, Y.Akaishi and H.Tanaka, {\sl Progr. Theor. Phys.} {\bf
79} (1988) 863.

\bibitem{7} 
C.Ciofi degli Atti, E.Pace and G.Salme, {\sl Phys. Rev.} {\bf C43}
(1991) 1155.

\bibitem{8} 
B.G.Zakharov, {\sl Sov. J. Nucl. Phys.} {\bf 38} (1983) 801.

\bibitem{9} 
L.G.Dakhno and N.N.Nikolaev, {\sl Nucl. Phys.} {\bf A436} (1985) 653.

\bibitem{10} 
R.I.Dzhibuti and R.Ya.Kezerashvili, {\sl Sov. J. Nucl. Phys.} {\bf 20}
(1974) 17.
\bibitem{11} 
G.D.Alkhazov, S.I.Belostotsky and A.A.Vorobyev, {\sl Phys. Rep.}
{\bf C42} (1978) 89.

\bibitem{12} 
T.Lasinski et al., {\sl Nucl. Phys.} {\bf B37} (1972) 1.

\bibitem{13} 
C.Lechanoine-LeLuc and F.Lehar, {\sl Rev. Mod. Phys.} {\bf 65}, 47
(1993).

\bibitem{14} 
A.Bianconi, S.Jeschonnek, N.N.Nikolaev and B.G.Zakharov, paper in
preparation.

\bibitem{15} 
N.N.Nikolaev, A.Szczurek, J.Speth, J.Wambach, B.G.Zakharov
and V.R.Zoller, {\sl Phys. Rev.} {\bf C50}, n.3 (1994) R1.

\bibitem{16} 
A.Kohama, K.Yazaki and R.Seki, {\sl Nucl. Phys.} {\bf A551 } (1993)
687.

\bibitem{17} 
N.N.Nikolaev, A.Szczurek, J.Speth, J.Wambach, B.G.Zakharov
and V.R.Zoller, {\sl Phys. Lett.} {\bf B317}, 281 (1993).

\bibitem{18} 
A.S.Rinat and B.K.Jennings, {\sl Nucl. Phys.} {\bf A568}, 873 (1994).

\bibitem{19} 
C.Ciofi degli Atti and S.Simula, {\sl Phys. Lett.} {\bf B325}, 276,
(1994).

\end{thebibliography}
\end{document}